\documentclass[%
aip,
rsi,
twocolumn,
amsmath,amssymb,
superscriptaddress,
 reprint,%
]{revtex4-1} 

\usepackage{amsmath,amssymb,amsfonts,bm}
\usepackage{graphicx,epstopdf}  
\usepackage{color}
\usepackage{extarrows}
\usepackage{enumitem}
\usepackage{empheq}
\usepackage{tikz}
\usetikzlibrary{matrix}

\allowdisplaybreaks[1]

\renewcommand{\d}{{\rm d}}
\newcommand{\ud}{{\rm d}}

\newcommand{\br}{{\bf r}}
\newcommand{\bx}{{\bf x}}

\newcommand{\la}{\langle}
\newcommand{\ra}{\rangle}

\newcommand{\Sec}[1]{Sec.\,\ref{#1}}
\newcommand{\nl}{\nonumber \\}
\newcommand{\be}{\begin{equation}}
\newcommand{\ee}{\end{equation}}
\newcommand{\bes}{\begin{equation*}}
\newcommand{\ees}{\end{equation*}}
\newcommand{\bsube}{\begin{subequations}}
\newcommand{\esube}{\end{subequations}}
\newcommand{\Eq}[1]{Eq.\,(\ref{#1})}
\newcommand{\Eqs}[1]{Eqs.\,(\ref{#1})}

\usepackage{tikz}
\usetikzlibrary{positioning, shapes.geometric}
\newcommand{\RN}[1]{%
  \textup{\uppercase\expandafter{\romannumeral#1}}%
}

\begin{document}

\title{Copula approach to exchange--correlation hole in many--electron systems with strong correlations}
\author{JingChun Wang}
\thanks{Authors of equal contributions}
\author{Yu Su}
\thanks{Authors of equal contributions}
\affiliation{
 Department of Chemical Physics,
 University of Science and Technology of China, Hefei, Anhui 230026, China}

\author{Haoyang Cheng}
\affiliation{
College of Electrical and Information Engineering,
Quzhou University,
Quzhou, Zhejiang 324000, China}

\author{Yao Wang}
\email{wy2010@ustc.edu.cn}
\author{Rui-Xue Xu}
\email{rxxu@ustc.edu.cn}
\affiliation{
 Department of Chemical Physics,
 University of Science and Technology of China, Hefei, Anhui 230026, China}

\date{\today}

 \begin{abstract}
Electronic correlation is a fundamental  topic in many--electron systems.
To characterize this correlation, one may introduce the concept of exchange--correlation hole.
In this paper, we first briefly revisit its definition and relation to electron and geminal densities, 
followed by their intimate relations to copula functions in probability theory and statistics.
We then propose a copula--based approach to estimate the exchange--correlation hole from the electron density.
It is anticipated that the proposed scheme would become a promising ingredient
towards the future development of strongly correlated electronic structure calculations.
\end{abstract}
\maketitle

\section{Introduction}
\label{thsec1}
Electronic correlation is a central topic in many--electron systems.
\cite{Sza96,Mar04,Xu07,Lev18}
It is rooted at the interaction between electrons and 
concerned with how the movement of one electron is influenced by the presence of all others.
For a non-relativistic many--electron system, the expectation value of electron--electron interaction,
$
\hat V_{ee}
$,
reads\cite{Xu07}
\be \label{hole_def}
\la \hat V_{ee}\ra=\frac{1}{2}\iint\!\d\br\d \br'\,\frac{\rho(\br)\rho(\br')+\rho(\br)h_{\rm xc}(\br,\br')}{|\br-\br'|}.
\ee
Here, $\rho(\br)$ and $h_{\rm xc}(\br,\br')$ are the electron density and exchange--correlation hole, respectively.
The latter plays a pivotal role in characterizing strongly correlated electronic structures. Exchange--correlation hole contributes the major part to the exchange--correlation energy functional, $E_{\rm xc}[\rho(\br)]$.
In density functional theory (DFT) calculations, the accuracy mostly depends  on
how to construct an approximate $E_{\rm xc}[\rho(\br)]$ of
good performances.
\cite{Hoh64B864,Koh65A1133,Par95,Bur12150901,Nar172315,Bec1418A301,Jon15897,Lan805469,Lan831809,Per868822,Lee872377,Per926671,Per9213244,Per963865,Per011,Per08136406}
Evidently, exchange--correlation hole
acquires its prominent position in further development of electronic structure theory with strong correlations.\cite{Noc917819,Coh12289,Sch172677}

We will see later that $h_{\rm xc}(\br,\br')$ is intimately related to the
copula function in probability theory and statistics. Copula, named by Sklar in 1959,\cite{Skl59229}  is defined as a class of multivariate cumulative distribution functions,
with the marginal properties all being  the uniform distributions in $[0, 1]$, referred as $U[0,1]$.
\cite{Skl59229,Nel07,Tri07}
 Copula--based methods
 are widely used in research fields,  including quantitative finance,\cite{Rod07401,Hu06717}  earthquake statistics,\cite{Nik08251,Li191950046}
  and cosmological data analysis.\cite{Sat11023501} 
  
  Naturally, it is also anticipated that copula is a powerful tool for the strongly correlated electronic structure theory.
In this paper, we propose a copula--based approach to estimate the exchange--correlation hole from the electron density.
It is exemplified with the Gaussian copula, following by discussions and some important issues toward the realistic simulations.
%
%

The remainder of the paper is arranged as
follows.
Section \ref{thsec2} comprises a brief introduction to the electron and geminal densities. We revisit their relations to the exchange-correlation hole in many--electron systems.
 In \Sec{thsec3}, we propose the copula scheme and develop its theoretical framework in detail.
The paper is finally summarized in \Sec{thsec4}.
%

\section{Reduced density matrices descriptions and exchange--correlation hole}\label{thsec2}

\subsection{Reduced density matrices}

For an $N$-electron system, the spinless reduced $p$-electron density matrix (R$p$DM), $\rho_p(\br_1\cdots\br_p,\br_1'\cdots\br_p')$, is defined as
\begin{align}
  \rho_p(\br_1\cdots\br_p,\br_1'&\cdots\br_p')\equiv 
  \begin{pmatrix}
    N\\p
  \end{pmatrix}
  \!\int\!\ud^N s\ud\br_{p+1}\cdots\ud\br_N  
  \nl 
  &\times\Psi(\br_1 s_1, \cdots,\br_p s_p,\bx_{p+1},\cdots,\bx_N)
  \nl 
  &\times\Psi^*(\br_1' s_1, \cdots,\br_p' s_p,\bx_{p+1},\cdots,\bx_N),
\end{align}
with the anti-symmetric electronic wave function, $\Psi(\bx_1, \cdots,\bx_N)\equiv \la \br_1 s_1, \cdots,\br_N s_N|\Psi\ra$ and $\bx_i\equiv \br_i s_i$. The electron density $\rho(\br)$ is therefore the diagonal element of R1DM, namely, 
\begin{align}\label{r1}
  \rho(\br) = \rho_1(\br,\br) \equiv Nf(\br).
\end{align}
We also define the geminal density, $\gamma(\br,\br')$, via the diagonal element of R2DM, as 
\begin{align}\label{r2}
  \gamma(\br,\br') \equiv \rho_2(\br\br',\br\br')\equiv {N\choose 2}f(\br,\br').
\end{align}
Here, $f(\br,\br')$ represents the joint probability density of finding electron $1$ at $\br$ and electron $2$ at $\br'$ simultaneously, while its marginal distribution
$
f(\br)= \int \!{\rm d}\br'\, f(\br,\br') 
$
 is the probability  of finding an electron at $\br$.
  If these two electrons are correlated, then the probability of finding electron $1$ at $\br$
  depends on the position $\br'$ of electron $2$, and vice versa.

To quantitatively describe the electronic correlation, we introduce the geminal correlation function, $\Pi(\br, \br')$, via
\be \label{eq1}
f(\br,\br')\equiv f(\br)f(\br')+f(\br)\Pi(\br, \br')f(\br').
\ee
 The geminal correlation function $\Pi(\br, \br')$
quantitatively represents this dependence,
\be
\Pi(\br, \br')=\frac{f(\br| \br')-f(\br)}{f(\br)}\ \ \text{with}\ \ f(\br|\br')\equiv \frac{f(\br, \br')}{f(\br')}.
\ee
Apparently, if these two electrons are uncorrelated, we have $f(\br|\br')=f(\br)$, leading to $\Pi(\br, \br')=0$.

\subsection{Exchange--correlation hole}

As known, the expectation value of the electron--electron interaction,
\begin{align}
\hat V_{ee}
=\frac{1}{2}\sum_{i\neq j}\frac{1}{|\hat \br_i-\hat \br_j|},
\end{align}
can be expressed as [{cf}.\,\Eqs{r2} and (\ref{eq1})]
\be\label{Vee}
\la \hat V_{ee}\ra=\iint\!\d\br\d \br'\,\,\frac{\gamma(\br,\br')}{|\br-\br'|}
\equiv
J+E_{\rm xc}.
\ee
 The Coulomb integral reads [{cf}.\,\Eq{r1}]
\begin{align}
 J = \frac{1}{2}\!
 \iint\!\d\br\d \br'\,\frac{\rho(\br)\rho(\br')}
 {|\br-\br'|}>0.
\end{align}
The exchange--correlation energy is [{cf}.\,\Eq{hole_def}]
\begin{align}
 E_{\rm xc}= \frac{1}{2}\!
 \iint\!\d\br\d \br'\,\frac{\rho(\br)h_{\rm xc}(\br,\br')}{|\br-\br'|}<0,
\end{align}
with
\be \label{pi2pi}
h_{\rm xc}(\br,\br')=[(N-1)\Pi(\br,\br')-1]f(\br').
\ee
This satisfies $\int\!\d\br'\,h_{\rm xc}(\br,\br') =-1$, the exchange--correlation hole theorem.\cite{Xu07} 

\subsection{Symmetry analysis}

Turn to the symmetry of $\Pi(\br,\br')$, on the basis of
group theory analysis,\cite{Dav76,Xu07,Bis93} with all the symmetry transformations forming a certain point group $G$.
For $\hat R \in G$, we have 
\be
\hat R\,\Pi(\br,\br')\equiv \Pi(\hat R ^{-1}\br,\hat R ^{-1}\br').
\ee
Consider then the basis wavefunctions of a
degenerate (not accidentally) energy eigenstate due to the symmetry.
The degenerate state $|\Psi\ra$ is usually a linear combination of a complete set of normalized orthogonal bases, $\{|\psi_k\ra\}$, as
\be
|\Psi\ra=\sum_{k=1}^{d}c_k|\psi_k\ra\ \ \text{with}\ \  \sum_{k=1}^{d}|c_k|^2=1.
\ee
Here, $d$ is the degree of degeneracy.
It is easy to show that
\be
\begin{split}
f(\br)&=\sum_{k,k'=1}^{d}c_k^{\ast}c_{k'}f_{kk'}(\br),
\\
f(\br,\br')&=\sum_{k,k'=1}^{d}c_k^{\ast}c_{k'}f_{kk'}(\br,\br'),
\end{split}
\ee
with
\be
f_{kk'}(\br)\equiv \int\!\d s \d\bx'\cdots\d\bx_N \,\psi_{k}^{\ast}(\{\bx_i\})\psi_{k'}(\{\bx_i\})
\ee
and
\be
f_{kk'}(\br,\br')
\!\equiv\!\int\!\d s\d s' \d\bx_3\!\cdots\!\d\bx_N
\psi_{k}^{\ast}(\{\bx_i\})\psi_{k'}\!(\{\bx_i\}).
\ee

According to the representation theory, the degenerate basis functions, $\{\psi_{k}; k=1,\cdots,d\}$, inside a $d$--dimensional irreducible representation $\Gamma$ of the point group $G$, transform only amongst themselves, i.e.,
\be
\hat R\,\psi_{k}=\sum_{k'} \psi_{k'}
\Gamma_{k'k}(\hat R);\ \ \hat R\in G,
\ee
 with $\Gamma_{k'k}(\hat R)$ being the  $d\times d$ representation matrix.
Since the wavefunction $\Psi$  belongs to $\Gamma$,  we know that
$
f_{kk'}(\br)$
belongs to $\Gamma\otimes \Gamma$, with
the transformation, 
\begin{align}
\hat Rf_{kk'}(\br)&=\int\!\d s \d\bx'\cdots\d\bx_N \,(\hat R\psi_{k})^{\ast}(\hat R\psi_{k'}),
\nl &
=\sum_{ij}\int\!\d s \d\bx'\cdots\d\bx_N \,\Gamma_{ik}^{\ast}(\hat R)\Gamma_{jk'}(\hat R)\psi_{i}^{\ast}\psi_j
\nl &
=\sum_{ij}f_{ij}(\br)\Gamma_{ik}^{\ast}(\hat R)\Gamma_{jk'}(\hat R).
\end{align}
According to the Uns\"{o}ld theorem, \cite{Xu07,Bis93}
this direct product representation, $\Gamma\otimes \Gamma$, contains precisely once the totally symmetric irreducible representation, in which
\be
f(\br)=
\frac{1}{d}\sum_{k=1}^{d}f_{kk}(\br),
\ee
retaining the symmetry.
Similarly, we know that
\be
f_{kk'}(\br,\br')\!=\int\!\d s\d s' \d\bx_3\cdots\d\bx_N \,\psi_k^{\ast}\psi_{k'}
\ee
also belongs to $\Gamma\otimes \Gamma$ with
\begin{align}
\hat Rf_{kk'}(\br,\br')
=\sum_{ij}f_{ij}(\br,\br')\Gamma_{ik}^{\ast}(\hat R)\Gamma_{jk'}(\hat R).
\end{align}
Therefore,
in the totally symmetric irreducible representation,
\be
f(\br,\br')=
\frac{1}{d}\sum_{k=1}^{d}f_{kk}(\br,\br')
\ee
is also invariant under the group point transformation.
Particularly, if a state is non-degenerate, belonging to a one--dimensional representation,
$f(\br)$ and $f(\br,\br')$ must be totally symmetric.

\section{Copula--based estimation scheme}\label{thsec3}

\subsection{Regular copula formulation}

Let us start with the general case of $n$ random variables with distribution function $f(x_1,\cdots,x_n)$.
The $n$-point cumulative density function (CDF) is defined as
\be \label{u}
F(q_1,\cdots,q_n)=\!\!\int_{-\infty}^{q_1}\!\d x_1 \cdots \!\int_{-\infty}^{q_n} \d x_n\, f(x_1,\cdots, x_n).
\ee
According to the Sklar's theorem, the CDF can be uniquely expressed as\cite{Skl59229,Nel07,Tri07}
\be \label{uu}
F(q_1,\cdots,q_n)=C(F_1(q_1),\cdots,F_n(q_n)).
\ee
Here
\be \label{uuu}
F_i(q_i)=\int_{-\infty}^{q_i}\!\d x_i\,f_{i}(x_i),
\ee
with $f_{i}(x)$ being the $i^{\rm th}$ marginal distribution density function that are positive.
Each function $F_{i}(q)$ is a nondecreasing function for the probability of the random variable less than or equal to $q$.
Apparently, $F_i(-\infty)=0$ and $F_i(+\infty)=1$.

In \Eq{uu}, $C(u_1,\cdots,u_n)$ is the copula function, associated with the CDF, and can be recast as
\be\label{Cdef}
C(u_1,\cdots,u_n)= F(F_1^{-1}(u_1),\cdots,F_n^{-1}(u_n)).
\ee
Taking derivative on both sides of \Eq{uu} with \Eq{u}, followed by the chain rule that deals with the differential of composite functions, we obtain
\be\label{fcff}
f(q_1,\cdots, q_n)=c(u_1,\cdots, u_n)\prod_{i=1}^{n}f_{i}(q_i).
\ee
Here
\be \label{ee}
c(u_1,\cdots, u_n)\equiv\frac{\partial C(u_1,\cdots, u_n)}{\partial u_1\cdots\partial u_n}.
\ee
As known $u_i=F_i(q_i)\in[0,1]$, thus 
\begin{align}
{\rm Prob}(u_i\leq b)&={\rm Prob}(F_i(q_i)\leq b)
\nl &
={\rm Prob}(F_i^{-1}(F_i(q_i))\leq F_i^{-1}(b))
\nl &
={\rm Prob}(q_i\leq F_i^{-1}(b))
\nl &
=F_{i}(F_i^{-1}(b))=b.
\end{align}
In other words, $u_i$ obeys $U[0,1]$, a uniform distribution on the interval $[0, 1]$. 

The copula function itself is a multivariate CDF whose all marginal distributions are $U[0,1]$. More specifically, $C(u_1,\cdots,u_n)$ maps $\{0\leq u_i\leq 1;i=1,\cdots,n\}$ to $[0,1]$, satisfying\cite{Nel07}
\be\label{property}
\begin{split}
  &C(1,\cdots,1,u_k,1,\cdots,1)=u_k;\\ 
  &C(u_1,\cdots,u_{k-1},u_k=0,u_{k+1}\cdots, u_n)=0;\\
  &0\leq \frac{\partial C(u_1,\cdots,u_n)}{\partial {u_k}}\leq 1.
\end{split}
\ee
Other intrinsic properties can be found in Ref.\,\onlinecite{Nel07}.

\subsection{Copula--based estimation scheme for exchange--correlation hole}
In the following, for bookkeeping we denote
\be  
\bm u\equiv(u_1,u_2,u_3)\equiv(F_{1}(q_1),F_{2}(q_2),F_{3}(q_3))
\ee
where $\{F_{i}(q_i)\}$ are the three marginal CDFs of $f(\br)\equiv \rho(\br)/N$.
From \Eq{fcff}, it is easy to see that
\be\label{x123}
f(\br)=c_3({\bm u})f_{1}({x})f_{2}({y})f_{3}({z}),
\ee
where
$
f_{1}(x)=\iint f(\br)
 \,\ud y\ud z
$; $f_2(y)$ and $f_3(z)$ are similar. 
All these CDFs depend on the 
parameters, as implied in \Eq{uuu}, and resulting
\be
c_3({\bm u})=\frac{f(\br)}{f_{1}(x)f_{2}(y)f_{3}(z)}.
\ee

For the joint distribution in \Eq{r2}, we have
\begin{align}
f(\br,\br')&=c_6(\bm u,{\bm u}')
f_{1}({x})f_{2}({y})f_{3}({z})
\nl &\quad\quad\quad\quad
\times f_{1}({x'})f_{2}({y'})f_{3}({z'}).
\end{align}
The composite copula function reads then [{cf}.\,\Eq{eq1}]
\begin{align}\label{r}
1+\Pi(\br,\br')=&\frac{f(\br,\br')}{f(\br)f(\br')}
=\frac{c_6(\bm u,{\bm u}')
}{c_3(\bm u)c_3(\bm u')}.
\end{align}
The exchange--correlation hole, \Eq{pi2pi}, becomes
\be 
h_{\rm xc}(\br,\br')=\bigg[(N-1)\frac{c_6(\bm u,{\bm u}')
}{c_3(\bm u)c_3(\bm u')}- N\bigg ]f(\br').
\ee

According to the Honhenberg--Kohn theorems, the exchange--correlation hole is a functional of density.
In copula--based approach, it is to
construct $c_6(\bm u,{\bm u}')$ from $f(\br)$.
The former is now assumed to be parameter--dependent and of the form $c_6(\bm u,{\bm u}';\{\theta_p\})$, whose integral correspondence is $C_6(\bm u,{\bm u}';\{\theta_p\})$; see \Eq{ee}.

For the present study, we set
\begin{align}\label{C6vv}
 C_6(\bm {u},\bm {u}')
=\int_{\bm{-\infty}}^{\bm\lambda}\int_{\bm{-\infty}}^{{\bm\lambda}'}\!\d\bm u'' \d\bm u'''\,\phi(\bm u'',{\bm u}''')
\end{align}
and exemplify it with the Gaussian copula via
\be 
\bm\lambda\equiv\Phi^{-1}(\bm{u})\equiv(\Phi_1^{-1}(u_1),\Phi_2^{-1}(u_2),\Phi_3^{-1}(u_3))
\ee
where
\be
\Phi_i(y)=\frac{1}{\sigma_i\sqrt{2\pi}}\int_{-\infty}^y\!\!\d y' \exp\Big[-\frac{(y'-a_i)^2}{2\sigma^2_i}\Big].
\ee
The corresponding integrand in \Eq{C6vv} is given by
\begin{align}\label{phiuu}
\phi(\bm u,\bm u')&=\sqrt{\frac{(\det {\bm \Sigma}_{uu})(\det {\bm \Sigma}_{u'u'})}{(2\pi)^6(\det{\bm \Sigma})\prod_{i=1}^3\sigma^2_i\sigma'^2_i}}
\nl &
\quad\times
e^{-\frac{1}{2}\big[(\bm u-\bm a)\cdot{\bm \Sigma}^{-1}_{uu'}\cdot(\bm u'-\bm a')+(\bm u'-\bm a')\cdot{\bm \Sigma}^{-1}_{u'u}\cdot(\bm u-\bm a)\big]}
\nl &
\quad\times
\prod_{i=1}^{3}e^{-(u_i-a_i)^2/(2\sigma^2_i)}e^{-(u'_i-a'_i)^2/(2\sigma'^2_i)},
\end{align}
with $\bm a=(a_1,a_2,a_3)$ and $\bm a'=(a'_1,a'_2,a'_3)$.  The covariance matrix is
\be
{\bm \Sigma}=\Biggl[\mkern-5mu
\begin{tikzpicture}[baseline=-.65ex]
\matrix[
  matrix of math nodes,
  column sep=1ex,
] (m)
{
{\bm \Sigma}_{uu} & {\bm \Sigma}_{uu'} \\
{\bm \Sigma}_{u'u} & {\bm \Sigma}_{u'u'} \\
};
\draw[dotted]
  ([xshift=0.5ex]m-1-1.north east) -- ([xshift=0.5ex]m-2-1.south east);
\draw[dotted]
  (m-1-1.south west) -- (m-1-2.south east);
\end{tikzpicture}\mkern-5mu
\Biggr].
\ee
In \Eq{phiuu}, $\{\sigma^2_i\}$ and $\{\sigma'^2_i\}$ are the diagonal elements of the submatrices
${\bm \Sigma}_{uu}$ and ${\bm \Sigma}_{u'u'}$, respectively.
Only the non-diagonal blocks, ${\bm \Sigma}_{uu'}$ and ${\bm \Sigma}_{u'u}$, are involved in the exponents.
This is
in parallel to the correlation between $f(\br)$ and $f(\br'\neq \br)$
without self--correlations.
The block diagonal parts, ${\bm \Sigma}_{uu}$ and ${\bm \Sigma}_{u'u'}$, only participate into the normalization.
The above $\bm a$, $\bm a'$, and ${\bm \Sigma}$ form the set of parameters, $\{\theta_p\}$.
In the present scenario, it is reasonable to set $\bm a=\bm a'$ and ${\bm\Sigma}_{uu}={\bm\Sigma_{u'u'}}$, due to the same function form of $f(\br)$ and $f(\br')$.

Finally, according to \Eq{ee}, the copula density reads
\begin{align}\label{Gaussian}
c_6(\bm u,{\bm u}')&=\phi(\bm \lambda,\bm \lambda')\prod_{i=1}^{3}
\frac{\d\Phi_i^{-1}(u_i)}{\d u_i}\frac{\d\Phi_i^{-1}(u'_i)}{\d u'_i}
\nl &=
\frac{\det {\bm \Sigma}_{uu}}{\sqrt{\det{\bm \Sigma}}}
e^{-(\bm \lambda-\bm a)\cdot{\bm \Sigma}^{-1}_{uu'}\cdot(\bm \lambda'-\bm a)}.
\end{align}
The parameters $\{\theta_p\}$ can be obtained directly from the data fitting. 
Upon obtaining $c_{6}({\bm u},{\bm u}')$, together with $c_{3}(\bm u)$ and $f(\br)$,
we can estimate the exchange--correlation hole, $h_{\rm xc}(\br,\br')$ [cf.\,\Eqs{pi2pi} and (\ref{r})],
and further the $\la\hat V_{ee}\ra$ via \Eq{Vee}.

To the end of this section, we would like to make some comments on the aforementioned copula--based scheme.
Firstly, the Gaussian copula [cf.\,\Eq{Gaussian}] is not the only choice.
Other commonly used copulas include $t$-copula, Gumbel copula, Frank copula, Clayton copula, and others. \cite{Skl59229,Nel07,Tri07}
Secondly, on the basis of the given copula, the task is to evaluate the involved parameters.
On the other hand, we may also use nonparametric estimation methods, via kernels, wavelets, splines, or neural networks. The hybrid usage of parametric and nonparametric method is also applicable. Last but not the least, the multivariate copula can be constructed via such as D-vines and canonical vines from lower order copulas.\cite{Joe11,Kra171} This will give rise to more detailed copula structures in relation to realistic molecules.

\section{Concluding remarks and prospect}\label{thsec4}
To conclude, in this paper we formulate a copula--based approach toward the 
accurate
exchange--correlation hole functional in strongly correlated systems.
%
%
%
Some issues would be important and addressed along this line, as follows.
The first one is to determine the asymptotic behaviours and physical constraints of the rigorous copula function.
This will reduce the functional space of the variation in the fitting process.
The second is to find out more symmetries beyond the point group,
especially those reduce the dimension of $\Pi(\br,\br')$ which is six at present.
A neutral network structure is also to be built for the copula function based on above considerations.
At last, a numerical recipe shall be developed
for the whole process,
to give the simulation results
and be compared with those from other quantum chemistry methods.

It is anticipated that the newly proposed scheme would become a promising ingredient
towards the future development of electronic structure calculations, including especially the DFT and its time--dependent  extension.\cite{Run84997,Mar04427,Cas12287} Generally speaking, the copula method would become an important tool in quantum chemistry simulations, just as 
its already wide use in other research fields.
\cite{Rod07401,Hu06717,Nik08251,Li191950046,Sat11023501} 

\begin{acknowledgements}
Support from the National Natural Science Foundation of China (Nos.\ 22103073 and 22173088)
is gratefully acknowledged.
The authors are indebted to YiJing Yan for valuable discussions.

\end{acknowledgements}



\end{document}